\documentclass[
    ,final            % use final for the camera ready runs
  ]
  {aipproc}

\layoutstyle{6x9}

\usepackage{amsbsy,amssymb,latexsym,amsmath}

\def\CF{{\cal F}}

\def\CL{{\cal L}}

\def\CV{{\cal V}}

\def\1{\mathbb I}
\def\im{{\rm Im}\,}

\def\CL{{\mathcal L}}
\def\CM{{\mathcal M}}

\def\CD{{\mathcal D}}

\def\CE{{\mathcal E}}
\def\CM{{\mathcal M}}

\def\CF{{\mathcal F}}

\def\Tr{{\rm Tr}}
\def\tr{{\rm tr}}

\def\CP{{\cal P}}

\newcommand{\bbR}{{\mathbb R}}

\newcommand\CW{{\mathcal W}}

\newcommand\CN{{\mathcal N}}
\newcommand\fR{{\mathfrak R}}

\newcommand\rangled{\rangle\! \rangle}
\newcommand\langled{\langle\!\langle}

\newcommand\bD{\mathbf{D}}

\begin{document}
\begin{flushright}{\footnotesize OCU-PHYS 256\\ 
OIQP-06-15}\end{flushright}\vspace{-10mm}

\title{Spontaneous Partial Breaking of $\CN = 2$ Supersymmetry
and the U(N) Gauge Model
\footnote{Talks 
given by HI
at 14th International Conference on Supersymmetry 
and the Unification of Fundamental Interactions (SUSY06), 
Irvine, California, USA, 12-17 June 2006, 
and given by KF at MSJ-IHES Joint Workshop on Noncommutativity, Bures-sur-Yvette, France, 
15-18 November 2006.
}
}


\author{K. Fujiwara}{
  address={Department of Mathematics and Physics,
Graduate School of Science,
Osaka City University
3-3-138, Sugimoto, Sumiyoshi-ku, Osaka, 558-8585, Japan}
}

\author{H. Itoyama}{
  address={Department of Mathematics and Physics,
Graduate School of Science,
Osaka City University
3-3-138, Sugimoto, Sumiyoshi-ku, Osaka, 558-8585, Japan}
}

\author{M. Sakaguchi}{
  address={Okayama Institute for Quantum Physics,
1-9-1 Kyoyama, Okayama 700-0015, Japan}
}

\begin{abstract}
We briefly review the properties of the $\CN=2$ $U(N)$ gauge model
with/without matters.
On the vacua,
$\mathcal{N}=2$ supersymmetry
and the gauge symmetry
are spontaneously
broken to $\mathcal{N}=1$ and a product gauge group,
respectively.
The masses of the supermultiplets appearing on the 
$\mathcal{N}=1$ vacua
are given.
We also discuss the relation to the matrix model.
\end{abstract}

\maketitle

%%%%%%%%%%%%%%%%%%%%%%%%%%%%%%%%%%%%%%%%%%%%
%% MAINMATTER
%%%%%%%%%%%%%%%%%%%%%%%%%%%%%%%%%%%%%%%%%%%%

%\section{Introduction}

The supercurrent algebra plays a key role in the 
partial breaking of global extended supersymmetries
\begin{eqnarray}
{\left\{ \bar{Q}_{\dot{\alpha}}^J,\mathcal{S}_{\alpha I}^m (x) \right\}=
2(\sigma^n)_{\alpha \dot{\alpha}} {\delta_I^J} T_n^m(x)
+(\sigma^m)_{\alpha \dot{\alpha}} }{C_I{}^J}
\label{supercurrent algebra}
\end{eqnarray}
where $\mathcal{S}_{\alpha I}^m$ are extended supercurrents,
$T_n^m$ is the stress-energy tensor 
and {$C_I{}^J$} is a field independent constant matrix,
which is permitted by the constraint
for the Jacobi identities~\cite{Lopuszanski:1978df}.
The new term does not modify the supersymmetry algebra on the fields.

Besides active researches on
the \textit{non-linear} realization of extended supersymmetry
in the partially broken phase,
Antoniadis, Partouche and Taylor (APT)
\cite{APT} (see also \cite{central_charge})
gave
a model
in which \textit{linearly} realized $\CN=2$ supersymmetry is
partially broken to $\CN=1$ spontaneously.
APT model is
$\mathcal{N}=2$ supersymmetric, self-interacting {$U(1)$} model
composed of one (or several) abelian constrained $\mathcal{N}=2$ vector multiplet(s) 
with electric \& magnetic Fayet-Iliopoulos (FI) terms.
In \cite{FIS1},
we have generalized this model to the {$U(N)$} gauge model
and shown that the $\CN=2$ supersymmetry is partially broken to
$\CN=1$ spontaneously.
Further in \cite{FIS3},
we have analyzed the vacua with broken gauge symmetry
and revealed the $\CN=1$ supermultiplets on the vacua.
The relation to the matrix model is discussed in \cite{F0609}.\footnote{
This series of works
\cite{FIS1,FIS3,F0609}
is based on $\CN=1$ superspace
and construction of the most general $\CN=2$
Lagrangian based on special K\"ahler geometry \cite{N=2 sugra}
which was developed after tensor calculus
\cite{N=2 sugra tensor}.
}
In addition,
a manifestly $\CN=2$ supersymmetric formulation of the
model coupled with/without $\CN=2$ hypermultiplets
was given in {\cite{FIS4}}
by using unconstrained $\CN=2$ superfields
on harmonic superspace
\cite{HS}.
These models 
contain the prepotential $\mathcal{F}$
as a prime ingredient.
So, our model should be regarded as a low-energy effective action
for systems given by $\CN=2$ bare actions spontaneously broken to $\CN=1$.
\smallskip

%%%%%%%%%%%%%%%%%%%%%%%%%%%%%%%%%%%%%%%%%%%%%%%%%%%%%%%%%%%%%%%%%%%%%%%%%%
%\section{$\CN=2$ $U(N)$ gauge model}
We introduce
$\CN=1$ superfields:
chiral $\Phi(x^m,\theta)=\Phi^a t_a \ni (A,\psi,F)$
and vector $V(x^m,\theta)=V^a t_a \ni (v_m,\lambda,D)$,
where 
$t_a$ ($a=0,...,N^2-1$)
generate $u(N)$ algebra,
$[t_a, t_b]=if^c_{ab}t_c$
and
$t_0$ generates the overall $u(1)$.
The model is composed of
the K\"ahler term
\begin{eqnarray}
\CL_{K}=
\int d^2\theta d^2\bar\theta ~K(\Phi^a,\Phi^{*a}),~~~~
 K=\frac{i}{2} (\Phi^a \CF^*_a
-\Phi^{{*a}} \CF_a),
\end{eqnarray}
where $\CF_a=\frac{\partial\CF}{\partial \Phi^a}$,
the $U(N)$ gauging counterterm $\CL_\Gamma$,
the $U(N)$ gauge kinetic term
\begin{eqnarray}
\CL_{\CW^2}=
-\frac{i}{4}\int d^2\theta ~ \CF_{ab}\,\CW^a\CW^b + c.c.~~~~~
{\mathcal W}_{\alpha}=-\frac{1}{4} \bar{D} \bar{D} e^{-V} D_{\alpha}
e^V={\mathcal W}_{\alpha}^a t_a~,
\end{eqnarray}
a gauge invariant superpotential term 
and the FI D-term
($e$ and $m$ are real constants)
\begin{eqnarray}
\CL_W=
 \int d^2\theta~ W(\Phi) +c.c.~,~~
\CL_{D}
=\xi \int d^2 \theta d^2 \bar{\theta} V^0
=\sqrt{2}\xi D^0~,~~~
W=eA^0+m\CF_0~.
\end{eqnarray}
In \cite{FIS1}, it is shown that the action
is invariant under $\fR$-action
which
 is composed of a discrete element of the $SU(2)$ R-symmetry
and a sign flip of the FI parameter
\begin{eqnarray}
R:\lambda_I^a\to \epsilon^{IJ}\lambda_J^a
~~~~~~\&~~~~~~~
R_\xi:
\xi \to -\xi~,
\end{eqnarray}
where
$\lambda_I^a=\bigl(\begin{smallmatrix} \lambda^a\\ \psi^a\end{smallmatrix}\bigr)$,
so that $S^{(+\xi)}\xrightarrow{R} S^{(-\xi)}\xrightarrow{R_\xi}S^{(+\xi)}$.
We have made the sign of the FI parameter manifest.
This ensures the $\CN=2$ supersymmetry of our action.
In fact,
acting $\fR$ on the first supersymmetry transformation
$\delta_{\eta_1}S^{(+\xi)}=0$, we have,
$
\delta_{\eta_1}S^{(+\xi)}=0
\xrightarrow{R}
R(\delta_{\eta_1})S^{(-\xi)}=0
\xrightarrow{R_\xi}
\fR(\delta_{\eta_1})S^{(+\xi)}=0
$,
which implies that
the resulting $\fR$-invariant action is invariant under the second supersymmetry
$\delta_{\eta_2}\equiv\fR (\delta_{\eta_1})$ as well.
By applying the $\fR$-action on the first supersymmetry transformation,
we obtain the $\CN=2$ supersymmetry transformation of the fermion as
\begin{eqnarray}
\boldsymbol{\delta \lambda}_J^{\ a} =
i(\boldsymbol{\tau} \cdot \widetilde\bD^a)_J{}^K \boldsymbol{\eta}_K
+\cdots~,~~~
\widetilde\bD^a =
-\sqrt{2} g^{ab^*}
\partial_{b^*}
\left( \boldsymbol{\mathcal{E}}A^{*0}+\boldsymbol{\mathcal{M}}
{\mathcal{F}}_0^* \right)
\end{eqnarray}
where 
$g_{ab}=\im \CF_{ab}$ and
$\boldsymbol{\tau}$ are Pauli matrices.
The rigid $SU(2)$ has been fixed
by making  $\boldsymbol{\mathcal{E}}$ and $\boldsymbol{\mathcal{M}}$ point 
to specific directions, $\boldsymbol{\mathcal{E}}=(0,\ -e,\ \xi)$ and 
$\boldsymbol{\mathcal{M}}=(0,\ -m,\ 0)$.
Under the symplectic transformation,
$\Omega=\bigl(\begin{smallmatrix}{A^0}\\{\CF_0}\end{smallmatrix}\bigr) \to \Lambda\Omega$,
$\Lambda\in Sp(2,\bbR)$,
$\bigl(\begin{smallmatrix} {-\CM}\\{\CE}\end{smallmatrix}\bigr)$ changes to
$\Lambda^{-1}\bigl(\begin{smallmatrix} {-\CM}\\{\CE}\end{smallmatrix}\bigr)$.
So the electric and magnetic charges are interchanged $(\CE',\CM')=(\CM,-\CE)$
when $\Lambda=\bigl(\begin{smallmatrix} 0   &-1    \\
      1 &0    
      \end{smallmatrix}\bigr)$.
This explains the name of the electric and magnetic FI terms.

%%%%%%%%%%%%%%%%%%%%%%%%%%%%%%%%%%%%%%%%%%%%%%%%%%%%%%%%%%%%%%%%%%

%\subsection{Analysis of vacua}
The vacua are specified by the scalar potential
given by
\begin{eqnarray}
\displaystyle \CV=
\frac{1}{8}g_{ab}
\CP^a\CP^b
+\frac{1}{2}g_{ab}
{\widetilde \bD}^a\cdot {\widetilde \bD}^{*b}
~,
~~
\CP^a\equiv g^{ab}\CP_b=-if^a_{bc}A^{*b}A^c~.
\end{eqnarray}
We find that for the case with
$\CF=\sum_{\ell=0}\frac{C_\ell}{\ell !}\Tr\Phi^\ell$,
the vacuum expectation values
are determined in the eigenvalue basis by
\begin{eqnarray}
\langled\CF_{\underline{i}\underline{i}}\rangled=
-2\left(\frac{e}{m}\pm i\frac{\xi}{m}\right)~.
\label{<F>}
\end{eqnarray}
The vev of $\CV$ is given by
$\langled\CV\rangled 
=2|m\xi|$.
On the other hand,
by applying supersymmetry transformation twice on the $U(1)_R$
charge conservation law \cite{Ferrara:1974pz},
we can read off the constant matrix
$C_I{}^J$ in (\ref{supercurrent algebra})
as $C_I{}^{J}=+4m\xi (\boldsymbol{\tau}_1)_I{}^{J}$.
Thus half of supercurrents
annihilates the vacuum.

\vspace{-2mm}
%\subsection{Gauge symmetry breaking}
Let $\langled A\rangled$
be 
$
\langled A\rangled=
{\rm diag}(\stackrel{N_1}{\overbrace{\lambda^{(1)},\cdots,\lambda^{(1)}}},
\stackrel{N_2}{\overbrace{\lambda^{(2)},\cdots,\lambda^{(2)}}},\cdots)
$ with 
$\sum_iN_i=N$
where $\lambda^{(k)}$ are complex roots of (\ref{<F>}),
then
$U(N)$ is broken to $\prod_{i} U(N_i$).
Unbroken $\displaystyle\Pi_i U(N_i$) is generated by
$t_\alpha\in \{t_a|[t_a,\langled A\rangled]=0\}$,
while broken ones by $t_\mu\in \{t_a|[t_a,\langled A\rangled]\neq 0\}$.
%\subsection{Partial supersymmetry breaking}
The $\CN=2$ supersymmetry is partially broken on the vacua
since
\begin{eqnarray}
\langled\frac{1}{\sqrt{2}}\boldsymbol{\delta}
(\lambda^{\underline{i}}\pm \psi^{\underline{i}})\rangled
=\pm im\sqrt{\frac{2}{N}}(\eta_1\mp\eta_2)~,
~~~
\langled\frac{1}{\sqrt{2}}\boldsymbol{\delta}
(\lambda^{\underline{i}}\mp \psi^{\underline{i}})\rangled
=0~,~~
\langled\boldsymbol{\delta \lambda}_I{}^{r}\rangled=0~.
\end{eqnarray} 
The mass spectrum is summarized as ($d_u\equiv \dim \prod_iU(N_i$))
\begin{eqnarray}
  \begin{array}{c|c|c|c}
 \mbox{field}      &\mbox{mass}    & \mbox{label}    &\mbox{\# of polarization states } \\
 \hline
 v_m^\alpha,~
\frac{1}{\sqrt{2}}(\lambda^\alpha\pm\psi^\alpha)       &0    &\mbox{A}    &2 d_u    \\
A^\alpha  ,~\frac{1}{\sqrt{2}}(\lambda^\alpha\mp\psi^\alpha)       
&| m\langled g^{\alpha\alpha}\CF_{0\alpha\alpha}\rangled |    &\mbox{B}    &2 d_u    \\
v_m^\mu+(T \overrightarrow{A^\mu})_2,~ 
\boldsymbol{\lambda}^\mu_I       &\frac{1}{\sqrt{2}}|f^{\tilde \mu}_{\mu \underline{i}}\lambda^{*\underline{i}}|    &\mbox{C}    &4(N^2-d_u)    
  \end{array}  \label{masses}
\end{eqnarray}
$\frac{1}{\sqrt{2}}
(\lambda^{0}\pm \psi^{0})$ is massless and thus is the Nambu-Goldstone (NG) fermion
for the partial breaking of $\CN=2$ supersymmetry to $\CN=1$.
Due to the $\CN=1$ supersymmetry on the vacua,
the modes
in (\ref{masses}) form $\CN=1$ multiplets:
(A)
massless $\CN=1$ vector multiples of spin($\frac{1}{2},1$),
(B)
massive $\CN=1$ chiral multiplets of spin($0,\frac{1}{2}$)
and (C)
two massive $\CN=1$ vector multiplets of spin($0,\frac{1}{2},1$).
%%%%%%%%%%%%%%%%%%%%%%%%%%%%%%%%%%%%%%%%%%%%%%%%%%%%%%%%%%%%%%%%%%%%%%%%%%%

\smallskip

%\section{Ralation to CDSW}
In \cite{FIS1}, it was shown that
the fermionic shift symmetry
arises in a certain limit of our model,
due to second supersymmetry,
which is spontaneously broken.
More precisely in \cite{F0609}, we study 
the decoupling limit of the NG fermion
and the fermionic shift symmetry in $\mathcal{N}=1 \ U(N)$ gauge model.
The fermionic shift symmetry plays a key role in the proof of the 
Dijkgraaf-Vafa conjecture which asserts that non-perturbative quantities in
$\mathcal{N}=1$ supersymmetric gauge theory can be computed by a matrix model.
The $\mathcal{N}=1$ action was obtained 
by ``softly" breaking of $\mathcal{N}=2$ supersymmetry
by adding the tree-level superpotential
$\int d^2 \theta \textrm{Tr} W(\Phi)$.
Thanks to the fermionic shift symmetry, effective superpotential is written as
$W_{\textrm{eff}}=\int d^2 \chi \mathcal{F}_p$ 
with some function $\mathcal{F}_p$,
which is
related to the free energy of the matrix model
\cite{Cachazo:2002ry}.

We consider the $\mathcal{N}=1$ $U(N)$ action
realized
on the vacua,
in which
the discrete $SU(2)$ R-symmetry
is broken by the superpotential.
How can we realize the situation in which
the NG fermion is decoupled
while
the superpotential remains non-trivial?
The answer is 
to consider the
large limit of the electric and magnetic FI parameters $(e, m, \xi)$. 
Parametrizing
\begin{eqnarray}
(e,\ m,\ \xi)= (\Lambda e',\ \Lambda m',\ \Lambda \xi')~,~~
\CF
=
\textrm{tr}
\left(
c_0 \textrm{\textbf{1}} +c_1 \Phi+\frac{c_2}{2} \Phi^2
\right)
+
\frac{1}{\Lambda} \sum_{\ell=3}^{n} \textrm{tr} \frac{c'_{\ell}}{\ell !} \Phi^{\ell}
\end{eqnarray}
and  taking the limit $\Lambda \rightarrow \infty$,
we derived
the general $\mathcal{N}=1$ action discussed in \cite{Cachazo:2002ry},
in which the NG fermion is 
decoupled while partial breaking of $\mathcal{N}=2$
supersymmetry is realized as before.
It shows that the fermionic shift symmetry is due to the free NG fermion. 
%%%%%%%%%%%%%%%%%%%%%%%%%%%%%%%%%%%%%%%%%%%%%%%%%%%%%%%%%%%%%%%%%%%%%%%%%%%
\smallskip

%%%%%%%%%%%%%%%%%%%%%%%%%%%%%%%%%%%%%%%%%%%%%%%%%%%%%%%%%%%%%%%%%%%%%
%\section{$\CN=2$ $U(N)$ gauge model  with matter} % $\CN=2$ hypermultiplets}

In \cite{FIS4}, we provide a manifestly $\CN=2$ supersymmetric
formulation of the $\CN=2$ $U(N)$ gauge model
with/without $\CN=2$ hypermultiplets.
It is convenient to use $\CN=2$ superfields in harmonic superspace:
vector multiplet
$V^{++a}(x^m, \theta^+,\bar\theta^+)
\ni (A^a,v_m^a,\lambda^{aI}_\alpha,D^{aIJ})$
and hypermultiplet $q^{+a}$ in the adjoint representation.
The action for the gauge field 
and the $U(N)$ gauged matter
is 
\begin{eqnarray}
S_{V+q}=-\frac{i}{4}\int d^4x\left[
(D)^4\CF(W)-(\bar D)^4\bar\CF(\bar W)
\right]
-\int dud\zeta^{(-4)}
\tilde q^+_a\CD^{++}q^{+a}
\end{eqnarray}
where $W$ is the curvature of $V^{++}$ and
$\CD^{++}q^{+a}=
D^{++}q^{+a}+iV^{++c}
if^a_{cb} q^{+b}$.
\newcommand\ad{{\rm ad}}
The electric FI term,
$
S_e
=\int dud\zeta^{(-4)}\tr(\Xi^{++}V^{++})+c.c.
$,
shifts the dual auxiliary field $D_D^{aIJ}$
in $W_D^a\equiv \CF_a$
by an imaginary constant.
So we introduce the magnetic FI term $S_m$
so as to shift the auxiliary field $D^{aIJ}$
as
$D^{aIJ}\to \bD^{aIJ}=D^{aIJ}+4i\xi^{IJ}\delta_0^a$.
By this, the $\CN=2$ supersymmetry transformation law
changes to 
$\delta_\eta\lambda^{aI}=(\bD^a)^I{}_J\eta^J+\cdots$,
under which
the total action
$
S=S_V+S^{\rm gauged}_q+S_e+S_m
$
is invariant.
We find that on the Coulomb phase
the $\CN=2$ supersymmetry is partially broken to $\CN=1$ spontaneously,
and
(\ref{<F>}) is reproduced by fixing $SU(2)$ appropriately.
A generalization to the case with $\CN=2$  local supersymmetry
is discussed in \cite{SUGRA PB}.

%%%%%%%%%%%%%%%%%%%%%%%%%%%%%%%%%%%%%%%%%%%%%%%%
%% BACKMATTER
%%%%%%%%%%%%%%%%%%%%%%%%%%%%%%%%%%%%%%%%%%%%%%%%

%\begin{theacknowledgments}
%This work is supported in part by the Grant-in-Aid for Scientific
%Research
%(No.16540262, No.17540262 and No.17540091) 
%from the Ministry of Education,
%Science and Culture, Japan.
%Support from the 21 century COE program
%``Constitution of wide-angle mathematical basis focused on knots"
%is gratefully appreciated.
%\end{theacknowledgments}

%%%%%%%%%%%%%%%%%%%%%%%%%%%%%%%%%%%%%%%%%%%%%%%%
%% The bibliography can be prepared using the BibTeX program or
%% manually.
%%
%% The code below assumes that BibTeX is used.  If the bibliography is
%% produced without BibTeX comment out the following lines and see the
%% aipguide.pdf for further information.
%%
%% For your convenience a manually coded example is appended
%% after the \end{document}
%%%%%%%%%%%%%%%%%%%%%%%%%%%%%%%%%%%%%%%%%%%%%%%%

%%%%%%%%%%%%%%%%%%%%%%%%%%%%%%%%%%%%%%%%%%%%%%%%
%% You may have to change the BibTeX style below, depending on your
%% setup or preferences.
%%
%%
%% For The AIP proceedings layouts use either
%%%%%%%%%%%%%%%%%%%%%%%%%%%%%%%%%%%%%%%%%%%%

\bibliographystyle{aipproc}   % if natbib is available
%\bibliographystyle{aipprocl} % if natbib is missing

%%%%%%%%%%%%%%%%%%%%%%%%%%%%%%%%%%%%%%%%%%%
%% You probably want to use your own bibtex database here
%%%%%%%%%%%%%%%%%%%%%%%%%%%%%%%%%%%%%%%%%%%
%\bibliography{bib-list}

\begin{thebibliography}{99}

\bibitem{Lopuszanski:1978df}
J.~T.~Lopuszanski,
%``The Spontaneously Broken Supersymmetry In Quantum Field Theory,''
Rept.\ Math.\ Phys.\  {\bf 13} (1978) 37.
%%
\bibitem{APT} 
I.~Antoniadis, H.~Partouche and T.~R.~Taylor,
%``Spontaneous Breaking of $\CN=2$ Global Supersymmetry,''
Phys.\ Lett.\ B {\bf 372} (1996) 83
[arXiv:hep-th/9512006].
%%

\bibitem{central_charge}
S.~Ferrara, L.~Girardello and M.~Porrati,
%``Spontaneous Breaking of N=2 to N=1 in Rigid and Local Supersymmetric
Theories,''
Phys.\ Lett.\ B {\bf 376} (1996) 275
[arXiv:hep-th/9512180].
%%
%%
\bibitem{FIS1}  K.~Fujiwara, H.~Itoyama and M.~Sakaguchi,
%``Supersymmetric $U(N)$ gauge model and partial breaking of $\CN = 2$
%supersymmetry,'' 
Prog.~Theor.~Phys. {\bf 133} (2005) 429
[arXiv:hep-th/0409060];~
%%
%\bibitem{FIS2}
%K.~Fujiwara, H.~Itoyama and M.~Sakaguchi,
%``$U(N)$ gauge model and partial breaking of $\CN = 2$ supersymmetry,''
in the proceedings of SUSY2004 [arXiv:hep-th/0410132].
%%
\bibitem{FIS3}
  K.~Fujiwara, H.~Itoyama and M.~Sakaguchi,
%  ``Partial breaking of $\CN = 2$ supersymmetry and of gauge symmetry in the $U(N)$
%  gauge model,''
  Nucl.\ Phys.\ B {\bf 723} (2005) 33
  [arXiv:hep-th/0503113].
%%
\bibitem{F0609}
  K.~Fujiwara,
%   ``Partial breaking of N = 2 supersymmetry and decoupling limit of
%  Nambu-Goldstone fermion in U(N) gauge model,''
  arXiv:hep-th/0609039.
%%

\bibitem{N=2 sugra}
L.~Andrianopoli, M.~Bertolini, A.~Ceresole, R.~D'Auria, S.~Ferrara and P.~Fre',
%``General Matter Coupled N=2 Supergravity,''
Nucl.\ Phys.\ B {\bf 476} (1996) 397
[arXiv:hep-th/9603004];
L.~Andrianopoli, M.~Bertolini, A.~Ceresole, R.~D'Auria, S.~Ferrara, P.~Fre and T.~Magri,
%``N = 2 supergravity and N = 2 super Yang-Mills theory on general scalar
%manifolds: Symplectic covariance, gaugings and the momentum map,''
J.\ Geom.\ Phys.\  {\bf 23} (1997) 111
[arXiv:hep-th/9605032].
%%
\bibitem{N=2 sugra tensor}
B.~de Wit, J.~W.~van Holten and A.~Van Proeyen,
%``Transformation Rules Of N=2 Supergravity Multiplets,''
Nucl.\ Phys.\ B {\bf 167} (1980) 186;
%%
J.~Bagger and E.~Witten,
%``Matter Couplings In N=2 Supergravity ,''
Nucl.\ Phys.\ B {\bf 222} (1983) 1;
%%
B.~de Wit, P.~G.~Lauwers, R.~Philippe, S.~Q.~Su and A.~Van Proeyen,
%``Gauge And Matter Fields Coupled To N=2 Supergravity,''
Phys.\ Lett.\ B {\bf 134} (1984) 37;
%%
B.~de Wit and A.~Van Proeyen,
%``Potentials And Symmetries Of General Gauged N=2 Supergravity - Yang-Mills
%Models,''
Nucl.\ Phys.\ B {\bf 245} (1984) 89;
%%
B.~de Wit, P.~G.~Lauwers and A.~Van Proeyen,
%``Lagrangians Of N=2 Supergravity - Matter Systems,''
Nucl.\ Phys.\ B {\bf 255} (1985) 569;
%%
E.~Cremmer, C.~Kounnas, A.~Van Proeyen, J.~P.~Derendinger, S.~Ferrara, B.~de Wit and L.~Girardello,
%``Vector Multiplets Coupled To N=2 Supergravity: Superhiggs Effect, Flat
%Potentials And Geometric Structure,''
Nucl.\ Phys.\ B {\bf 250} (1985) 385;
%%
H.~Itoyama, L.~D.~McLerran, T.~R.~Taylor and J.~J.~van der Bij,
%``N=2 No Scale Supergravity,''
Nucl.\ Phys.\ B {\bf 279} (1987) 380;
%%
R.~D'Auria, S.~Ferrara and P.~Fre,
%``Special And Quaternionic Isometries: General Couplings In N=2 Supergravity
%And The Scalar Potential,''
Nucl.\ Phys.\ B {\bf 359} (1991) 705.

%%

\bibitem{FIS4}
  K.~Fujiwara, H.~Itoyama and M.~Sakaguchi,
%   ``Partial supersymmetry breaking and $\CN = 2$ $U(N_c)$ gauge model with
%  hypermultiplets in harmonic superspace,''
  Nucl.\ Phys.\ B {\bf 740} (2006) 58
  [arXiv:hep-th/0510255];~
% K.~Fujiwara, H.~Itoyama and M.~Sakaguchi,
%   ``Supersymmetric $U(N)$ gauge model and partial breaking of $\CN = 2$
%  supersymmetry,''
 in the proceedings of ``Frontier of Quantum Physics'' [arXiv:hep-th/0602267].
  
%%
\bibitem{HS}
  A.~S.~Galperin, E.~A.~Ivanov, V.~I.~Ogievetsky and E.~S.~Sokatchev,
  ``Harmonic superspace,'' Cambridge University Press, 2001.

%%
\bibitem{Cachazo:2002ry}
F.~Cachazo, M.~R.~Douglas, N.~Seiberg and E.~Witten,
%``Chiral rings and anomalies in supersymmetric gauge theory,''
JHEP {\bf 0212} (2002) 071 
[arXiv:hep-th/0211170].

\bibitem{Ferrara:1974pz}
S.~Ferrara and B.~Zumino,
%``Transformation Properties Of The Supercurrent,''
Nucl.\ Phys.\ B {\bf 87}, 207 (1975);
%%
%\bibitem{Itoyama:1996hh}
H.~Itoyama, M.~Koike and H.~Takashino,
%``N = 2 supermultiplet of currents and anomalous transformations in
%supersymmetric gauge theory,''
Mod.\ Phys.\ Lett.\ A {\bf 13}, 1063 (1998)
[arXiv:hep-th/9610228].

\bibitem{SUGRA PB}
  H.~Itoyama and K.~Maruyoshi,
%   ``U(N) gauged N = 2 supergravity and partial breaking of local N = 2
  %supersymmetry,''
  arXiv:hep-th/0603180;
  K.~Maruyoshi,
%``Gauged N = 2 supergravity and partial breaking of extended supersymmetry,''
  arXiv:hep-th/0607047.






\end{thebibliography}

%%%%%%%%%%%%%%%%%%%%%%%%%%%%%%%%%%%%%%%%%%%
%% Just a reminder that you may have to run bibtex
%% All of it up to \end{document} can be removed
%% if you don't like the warning.
%%%%%%%%%%%%%%%%%%%%%%%%%%%%%%%%%%%%%%%%%%%
\IfFileExists{\jobname.bbl}{}
 {\typeout{}
  \typeout{******************************************}
  \typeout{** Please run "bibtex \jobname" to optain}
  \typeout{** the bibliography and then re-run LaTeX}
  \typeout{** twice to fix the references!}
  \typeout{******************************************}
  \typeout{}
 }

%\section{refs}

\end{document}